\documentstyle[10pt,a4,twocolumn,epsf]{article}
\begin{document}

\twocolumn[

\

\

\

\

\

\

\

\

\begin{center}
{\large The Quantum Hall Effect in quasi-1D
conductors}
\end{center}

\small{

\begin{center}

  G. Montambaux  and D. Zanchi
\medskip

{\it Laboratoire de Physique des Solides,  associ\'e au
CNRS \\ Universit\'{e} Paris--Sud \\ 91405 Orsay, France}
\end{center}
\bigskip

\noindent
---------------------------------------------------------------------------------------------------------------------------------------------------------------

{\bf Abstract}
\medskip

The theory and experiments showing Quantum Hall effect in the
quasi-one-dimensional conductors of the Bechgaard salts family are briefly
reviewed.
The sign reversals observed under some  experimental conditions
are explained
within the framework of the Quantized Nesting Model. The sequence of
reversals is driven by slight modifications of the geometry of the Fermi
surface. It is explained why only even phases can have sign reversals and
why negative phases are less stable than positive ones.
\medskip

{\it keywords:} Many-body and quasiparticle theories,
Hall effect, Magnetotransport, Magnetic phase transitions, Organic
conductors, Bechgaard salts

\noindent
---------------------------------------------------------------------------------------------------------------------------------------------------------------
}
\vspace{1cm}

]

\small{

{\bf 1. Introduction}
\medskip

The organic conductors of the Bechgaard salts family
( $(TMTSF)_2X$ where $TMTSF$ $=$ tetramethyselenafulvalene) have remarkable
properties in a magnetic field. Although these compounds are metals with a
large number of carriers ${\cal{N}}$, they exhibit Quantum Hall Effect
which normally would require a small filling factor $\nu = {\cal{N}}h/e H$.
Moreover, the Fermi surface  of these systems is open, made of two
almost planar sheets so that no orbital quantization is
expected\cite{Review}.

The quantization of Hall effect in these materials results
from the magnetic field induced low temperature instability of the metallic
phase versus the
formation of an ordered state in which a gap opens close to the Fermi level.
As a result, the ordered phase (a Spin Density Wave state) contains
a much smaller number of carriers above (electrons) or below (holes) the
Fermi level. These carriers form closed pockets which are quantized into
Landau levels, giving rise to the Quantum Hall Effect\cite{Montambaux91}.

The Bechgaard
 salts      are
strongly  anisotropic systems, with a typical hierarchy of transfer
integrals: $t_a\simeq 3000K, t_b\simeq 300K, t_c \simeq 10K$. In three
members of this family  ($X = ClO_4, PF_6, ReO_4$), the metallic phase is
 destroyed by a moderate magnetic field $H$ applied along
the $\bf c^*$ direction, perpendicular to the most
 conducting planes $(\bf a,\bf b)$. The Field Induced
 phase consists in a series of Spin Density Wave (FISDW)
 subphases, separated by first order transitions\cite{Cooper89,Hannahs89}.
This field induced cascade of quantized phases results from an interplay
between the nesting properties of the Fermi surface and the quantization
of electronic orbits in the field: the wave vector of the SDW adjusts
itself with the field so that unpaired carriers in the subphases always
fill an integer number of Landau levels. As a result, the
number of carriers in each subphase is quantized and so is the Hall effect:
$\sigma_{xy}= 2 N e^2/h$ (a factor
 $2$ accounts for spin degeneracy)\cite{Poilblanc86,Yakovenko91}.
The apparition of these phases results from a new structure of the metallic
phase in a field: because of the Lorentz force, the electronic
motion becomes periodic and confined along the direction of the
chains of high conductivity ($\bf a$ direction). As a result of this
effective reduction of dimensionality, the metallic phase becomes
unstable\cite{Model1}. In addition, the electrons
experience a periodic motion in real space, characterized by the
wave vector $G= e H b/\hbar$, $b$ being the interchain distance.
Consequently,
the spin susceptibility, instead of having one logarithmic divergence at
 $2k_F$, exhibits a series of divergences at quantized values of the
longitudinal component of the wave
vector $Q^n_\parallel =
2 k_F + n G$\cite{Montambaux91,Model2,Montambaux85b}.
 The largest divergence signals the appearance of a SDW
phase with quantized  vector $Q_\parallel = 2 k_F + N G$.
These ideas have been formalized in the so-called Quantized Nesting
Model  which describes most of the features of the
phase diagram in a magnetic field, in particular the observed Hall
plateaus\cite{Montambaux91,Model2}.

In this paper, we review the main features of the Quantized Nesting
Model which  describes the observed cascade of FISDW transitions. Then we
generalize this model  to explain the change in sign in the Hall plateaus
which is observed in the salts $X=PF_6, ClO_4, ReO_4$ under certain
conditions\cite{Zanchi96}. \bigskip

{\bf 2. The Quantized Nesting Model}
\medskip

The model which describes the FISDW starts from the effective Hamiltonian
for the metallic phase: \begin{equation}
{\cal H}=v_F(|k_x|-k_F)+t_\perp(k_y b)
\end{equation}
$t_\perp(k_y b)$ is a periodic function which
 describes a warped Fermi surface. It satisfies
the properties $ t_\perp(p+ 2 \pi) = t_\perp(p)$ and $t_\perp(-p) =
t_\perp(p)$. More specifically the following function has been chosen:
\begin{equation}   \label{dispersion1}
t_\perp(p)=-2t_b\cos p-2t_b^{\prime }\cos 2p
\end{equation}
Although it is essential to explain the existence of a threshold field
for the cascade of FISDW, the coupling in the third direction is omitted
since it is known that
it does not play an important role in the sequence of
subphases\cite{Montambaux91,Montambaux85a}.
This dispersion relation contains the minimum number of parameter to
describe the FISDW cascade.
The first harmonics ($t_b$) of the dispersion along the transverse direction
$y$
describes the warping of the FS with a perfect nesting at wave vector $(2
k_F, \pi/b)$. The second harmonics ($t'_b$) induces a deviation
from perfect nesting which leaves a small number of carriers quantized
into Landau bands. Its amplitude fixes the period of the
cascade\cite{Montambaux91,Model2,Montambaux85b}.
Typically in Bechgaard salts, $t'_b \simeq 10
K$.
This term may have two origins. One
is  the linearization
of the dispersion relation along the $x$ direction and is given by
$t_b^{\prime}=-(\cos k_Fa/4\sin {}^2k_Fa)t_b^2/t_a$\cite{Yamaji82}. Since
the band is 3/4
filled, $k_Fa=3\pi /4a$ and $t_b^{\prime }$ is  positive.
Other contributions may result directly from next nearest neighbor
coupling\cite{Yamaji86}.
The instability of the metallic phase can be described by  the spin
susceptibility $\chi_0$ whose maximum gives access to the wave vector
of the ordered phase.
It is given by \begin{equation}
\chi_0({\bf Q}) = \sum_n I_n^2(Q_\perp) \chi_0^{1D}(Q_\parallel - n G)
\end{equation}
This forms exhibits the variation
of $\chi_0$ as the sum of one-dimensional
 contributions  $\chi_0^{1D}$ shifted from the magnetic
field wave vector $G= e H b / \hbar$\cite{Montambaux86}.
The $I_n$ depend on the zero field dispersion relation\cite{Montambaux86}.
\begin{equation} \label{In}
I_n(Q_\perp) = \langle
e^{i[T_\perp(p+Q_\perp/2)+T_\perp(p-Q_\perp/2)+n
p]} \rangle \end{equation}
where $T_\perp(p)=(2/\hbar \omega_c) \int_0^p  t_\perp(p') dp'$ and
 $\langle ... \rangle$ is the average over $p$. $\omega_c = e v_F b H/2$
is the cyclotron frequency of the open periodic motion in the metallic phase
and $\hbar \omega_c$ is the separation between Landau bands in the FISDW
phases. $\chi_0$ has logarithmic divergences
at quantized values of the wave vector  and the largest divergence
signal the formation of a FISDW at the corresponding wave vector
 ${\bf Q}_N = (Q_\parallel, Q_\perp)=(2 k_F + N eHb/\hbar,Q_\perp)$.
When the field is varied, each of the peaks becomes in turn the absolute
maximum. When $H$ decreases $N$ increases {\it monotonously}, $N=0
\rightarrow 1 \rightarrow 2 \rightarrow 3 \rightarrow 4 \rightarrow 5 ...$.

\begin{figure}[htb]
\centerline{ \epsfysize 7cm
\epsffile{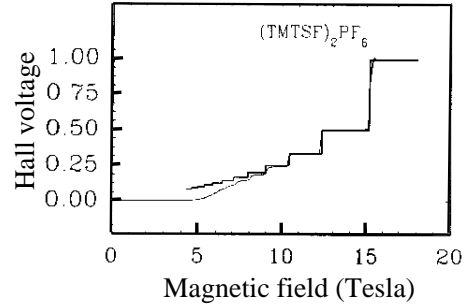}}
\caption{{\footnotesize The quantum Hall sequence in $(TMTSF)_2PF_6$
\protect\cite{Balicas95} compared
with the result of the Quantized Nesting Model, for
$t_b=300K$,$t'_b\simeq 20K$. At low field, the temperature, $150mK$ is too
large to allow for the formation of plateaus and the Hall effect is smaller
than the quantized value. Below $5T$, the FISDW sequence is destroyed due
to the coupling in the third direction.} } \label{fig0} \end{figure}

It is possible to go beyond the description of the transition line and to
give a complete description of the SDW
subphases\cite{Yamaji85,Virosztek86,Montambaux88a}. The divergence of the
susceptibility  $\chi = \chi_0 /(1-\lambda \chi_0)$ at a quantized wave
vector signals the spontaneous formation of  a density wave with this wave
 vector.
This new periodicity couples the eigenstates of the metallic phase and opens
a       {\it series} of gaps at quantized values of the wave vector. The
total energy is maximum if the largest of these gaps stays at the Fermi
level. This leads to the quantization of the Hall effect and vanishing
dissipation.

The Hall conductivity can be calculated from Streda
formula\cite{Poilblanc86,Streda82}
\begin{equation}
\sigma_{xy} = -e ({\partial {\cal{N}}    \over \partial
H})_{T,\mu,\bf Q}
\end{equation}
where ${\cal{N}}$ is
 the carrier density. The derivative is taken {\it at fixed external
parameters, in particular at fixed wave vector \bf Q}. Since
\begin{equation}
{\cal{N}} ={2\over 4 \pi^2} {2 \pi \over b} 2 k_F =2  {1 \over 2 \pi b}
(Q_\parallel  - N {e H b \over
\hbar})
\end{equation}
$\sigma_{xy}$ can be rewritten as:
\begin{equation}
\sigma_{xy}= 2 N {e^2 \over h}
\end{equation}
The additional factor $2$ accounts for spin degeneracy.
The quantization of the Hall effect has also been proven by
Yakovenko by a direct derivation from the Kubo formula\cite{Yakovenko91}.
This model, with the above dispersion relation(\ref{dispersion1}),
describes remarkably well the phase diagram and the QHE in
$(TMTSF)_2PF_6$,
in which there is a remarkable quantization of the Hall effect. In
ref.\cite{Balicas95}, the
transition fields $H_n$ obey the relation $H_n = H_f / (n+\gamma)$ with $H_f
\simeq 67T$ and $\gamma \simeq 3.5$. This is in excellent agreement with the
prediction of the nesting model (fig. \ref{fig0}), where one finds $e v_F
H_n b \simeq
5.8 t'_b/(n+\gamma)$ with $\gamma \simeq 3.45$. Using the parameters of
the Bechgaard salts, $ev_F b \simeq 1.67 K/T$, a value $t'_b \simeq 19.5K$
fits the observed cascade  ($H_f$ depends strongly on $t'_b$
and thus on pressure: it varies between different
experiments\cite{Cooper89,Hannahs89,Balicas95}).

The nesting model also describes very well the thermodynamic properties (
magnetization, specific heat) of the FISDW phases in the $PF_6$ and $ClO_4$
salts\cite{Montambaux91}.
\bigskip

{\bf 3. The "Negative" phases}
\medskip

However, one of the most puzzling  unexplained experimental results is
certainly the possibility of a reversal of
the Hall effect when the field varies: although most of the phases exhibit
the same sign of the Hall voltage (by convention we will refer to these
plateaus as the {\it positive} ones),
 it has been discovered by Ribault that  {\it negative}
plateaus may appear in $(TMTSF)_2ClO_4$  under certain conditions of
cooling rate\cite{Ribault85}.
Such negative plateaus  have been reproduced and also
found in $(TMTSF)_2PF_6$, where their existence crucially depends on the
pressure\cite{Cooper89,Balicas95,Ribault85,Piveteau86}.
 One of the puzzling aspect of these negative plateaus is that most often
they resemble a dip rather than a plateau and they seem less stable than
 the positive ones.

Quite recently,
  by a conditioning procedure in which current pulses depin the FISDW
from lattice defects and tend to reduce hysteresis,
 it has been shown  unambiguously that
there exists at least one phase characterized by a well-formed negative
plateau with a quantized
value of $\sigma_{xy}$ ($N=-2$)\cite{Balicas95}. In this experiment, the
sequence
of  plateaus  obtained by decreasing the field can be clearly identified
with the quantum numbers $N=1,2,-2,3,4,5,6,7$.
 Although there is only one negative plateau in this experiment, an older
work have shown  a sequence of phases which could be labeled
by $N=1,2,-2,4,-4,5,6$\cite{Cooper89}. Note that the negative plateaus
are  labeled by {\it even} numbers only. In others salts, $ClO_4$ and
$ReO_4$,
there are also several negative features but it is more difficult to ascribe
them a well defined quantum number\cite{Ribault85,Kang91} (
$ReO_4$: $1,2,-2,?$). Moreover, in these two materials, the
situation is complicated by the anion ordering which certainly affects the
apparition of subphases\cite{Osada92}. In $ClO_4$, the existence of
negative phases is also very dependent on pressure\cite{Kang93}.

One very important key is to notice that at least the phase
$N=-2$ extends up to the
metallic phase\cite{Balicas95}. This feature cannot be explained by a theory
based on multiple order
parameter states which would appear only at low temperature\cite{Machida94}.
 On the contrary, a description of the metallic
 phase instability {\it should}  explain the existence of such phases with
an appropriate form of the Fermi surface.
In the next section, we show that
 an appropriate and very slight modification of the
Fermi Surface (FS)  can lead to the
sequence of plateaus $1,2,-2,3,4,-4,5$. Our result strengthens the
validity  of the Quantized Nesting Model used to describe
the phase  diagram of Bechgaard salts in a magnetic field.
\bigskip

{\bf 4. Sign reversals of the Quantum Hall Effect}
\medskip

This change in sign in the Hall effect cannot be explained with the model
using the above dispersion relation(\ref{dispersion1}). The reason is the
following:
when the field
varies, the nesting
vector oscillates around its zero field value, which connects the inflexion
points of the Fermi surface\cite{Montambaux85a}. The sign of the carriers is
thus given by
the position of the inflexion point. A simple geometric analysis shows that
$sign(N)=sign(Q_\parallel - 2 k_F) =  sign(t'_b)$,
and one sees on figs. \ref{fig1}a,b. that
$\chi_0({\bf Q}_N) > \chi_0({\bf Q}_{-N})$
for all magnetic fields, so that the sign reversal could not be explained in
this framework.

 \begin{figure}[htb]
 \centerline{ \epsfxsize 6cm
\epsffile{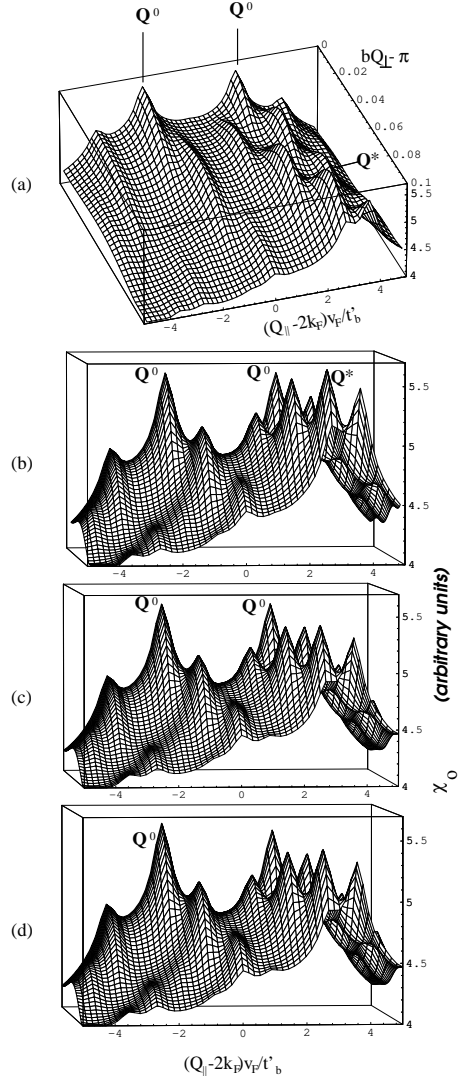}}
 \caption{
\footnotesize{a) $\chi_0({\bf Q})$ in a finite magnetic field for the standard
model, $t'_b=10K,t_3=t_4=0$. The best nesting vector is ${\bf Q}^*$. ${\bf
Q}^0$
is a degenerate secondary maximum. b) Same parameters. c) A finite $t_3=
10K$
alters the best nesting and  ${\bf Q}^0$ is now the degenerate best nesting
vector. d)  A finite $t_4= 0.2K$ lifts the degeneracy,
 leading to a negative quantum number.} } \label{fig1} \end{figure}

However, the fact that the sign of the Hall sequence is fixed
cannot be  a general
feature. The detailed structure of $\chi _0$ has to depend on the fine
geometry of the FS\cite{Montambaux85a}. If two regions of the Fermi surface
exhibit almost equally good nesting properties, one can imagine that the SDW
vector will oscillate between positive  ($Q_\parallel > 2 k_F$) and
negative ($Q_\parallel < 2 k_F$)  regions.
 In this paper, we show  that, by a small change in the dispersion relation,
these sign reversals can be described
as an equilibrium solution of the Nesting Model.

First, note  that there are several maxima
for each value of the quantum number and that
there is a secondary maximum, noted $\bf Q^0$, on the $Q_\perp = \pi / b$
line, but  {\it for even phases only} . This is
because $I_{2M+1}(\pi / b) =0$, a property which can be
checked directly
from eq. (\ref{In}). This property has a simple semiclassical qualitative
explanation: when  $Q_\perp = \pi / b$, the pocket of unnested carriers
is be splitted into {\it two} pockets of equal size. Quantization in each of
these two pockets implies an {\it even} quantization of the total number of
unnested carriers. For the standard
model, the absolute maximum, noted $\bf Q^*$, always lies {\it outside} the
$Q_\perp = \pi
/ b$ line\cite{Montambaux85b,Montambaux85a}. This is a reminiscence of the
position of the zero field best nesting
vector\cite{Montambaux85a}(figs.\ref{fig1},\ref{fig2}).

 \begin{figure}[htb]
\centerline{ \epsfxsize 6cm \epsffile{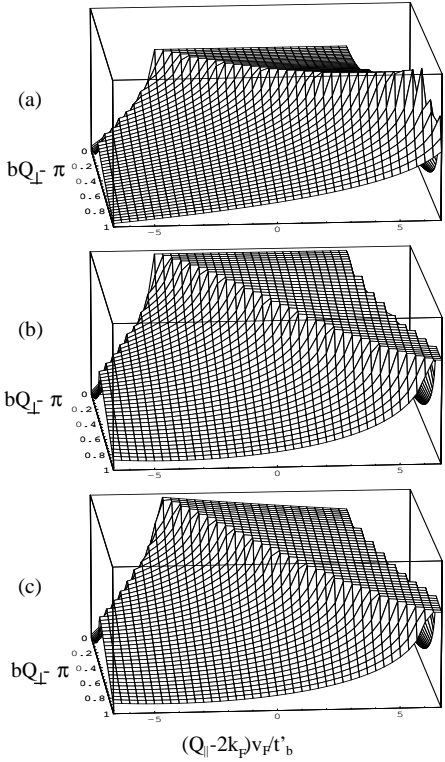}} \caption{\footnotesize{
a) $\chi_0({\bf Q})$ in zero field for the
standard model. There
is a  maximum corresponding to the inflexion point of the FS
with $Q_\parallel > 2k_F$.
b) When $t_3 \neq 0$,  the maximum is moved along  the degenerate $Q_\perp
= \pi /b$ line. c) When  $t_4 \neq 0$, the maximum has $Q_\parallel <
2k_F$. To increase the effect on the figures, we have chosen large values
of the parameters $t'_b=60K$, $t_3=20K$, $t_4=2K$.
} } \label{fig2}
\end{figure}

In order to change the geometry of the Fermi surface and to change the
nesting at the inflexion point, we propose that next harmonics in the
dispersion relation
may play a very important role. The dispersion relation is now taken as:
\begin{equation} \label{dispersion2}
t_\perp(p)=-2t_b\cos p-2t_b^{\prime }\cos 2p-2t_3 \cos 3p - 2 t_4 \cos 4p
\end{equation}

In Bechgaard salts, the additional harmonics exist and
 result directly from next nearest neighbor coupling\cite{Yamaji86}.
A very slight
modification of the FS induced by a third ($t_3$) and a fourth
($t_4$) harmonics in the transverse
direction is enough to explain the existence of new phases with a change in
the sign of the Hall effect.   We explain now why these two terms are
equally important to describe the sign reversals.

\begin{figure}[htb]
\centerline{
\epsfxsize 6cm
\epsffile{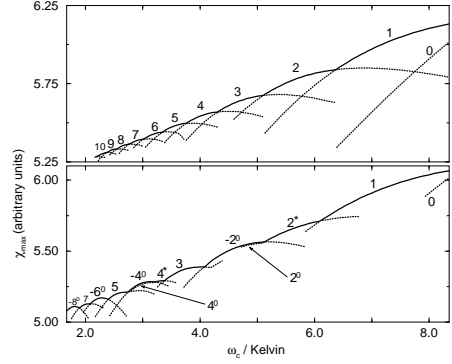}}
\caption{\footnotesize{ $\chi_{max}(H)$ for  $t_3 = t_4=  0$ and for  $t_3
\neq 0$,   $t_4 \neq 0$.}
}
\label{fig4}
\end{figure}

\begin{figure}[bht]
\centerline{
\epsfxsize 6cm
\epsffile{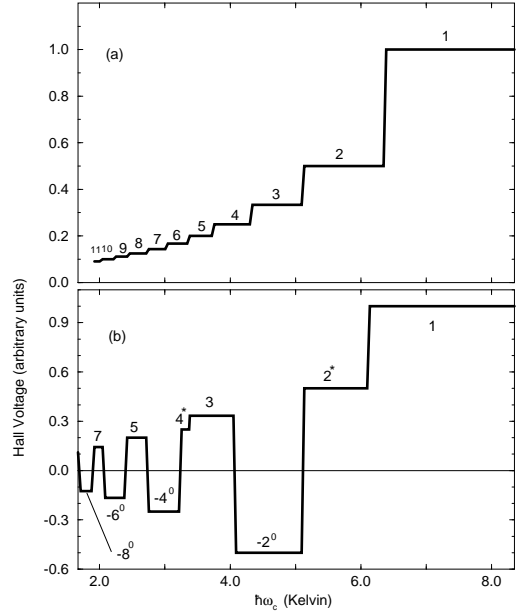}}
\caption{ \footnotesize{ Hall voltage versus field for a)  $t'_b = 10K$,
$t_3 = t_4=0$   b) $t_3 = 7K$,
  $t_4 =.025K$, obtained from $\chi_0(\bf Q)$ at $T=0.5K$.}
}
\label{fig3}
\end{figure}

 The effect of a third harmonics $t_3$ in the  dispersion
relation is to deteriorate the nesting at the inflexion point. This is seen
on figs \ref{fig2} for zero field. As a result, when the field is applied,
the odd
absolute maxima  have  still  $Q_\parallel >2 k_F$ but the even maxima can be
of two different natures depending on the field:  either they stand on the
$Q_\perp = \pi / b$ line ($Q^0$ on figs. \ref{fig1},\ref{fig2}) or
towards the zero field best
nesting vector ($Q^*$ on fig. \ref{fig1}). When they lie on the $Q_\perp =
\pi /
b$ line, these  maxima are degenerate: $\chi_0({\bf Q}^0_{2M} = \chi_0({\bf
Q}^0_{-2M})$. This degeneracy had already been noticed in the
past\cite{Montambaux88b,Machida94}.   It would lead to a phase diagram where
$-2$ and $2$ are degenerate, which is not the case.

 We have found
that this degeneracy can  be removed by the addition of a fourth harmonics
of amplitude $t_4$. On the  $Q_\perp = \pi / b$ line, the $I_N$ are given
by: \begin{equation}
I_{N}(\pi / b) = \langle \exp [ {4 i  \over \hbar \omega_c} (t'_b \sin 2p -
{t_4 \over 2} \sin 4p) +  i N p] \rangle
\end{equation}
By changing $N$ into $-N$ and $p$ into $p + \pi /2$ ,
one has:
\begin{equation}
I_{-N}(\pi / b) = (-1)^{N/2}\langle \exp [ {4 i  \over \hbar \omega_c} (t'_b
\sin 2p + {t_4 \over 2} \sin 4p) +  i N p] \rangle
\end{equation}
One immediately sees that if $N$ is odd, $I_N=0$ as stated above. If $N$ is
even, the degeneracy is broken by a non-zero $t_4$.
  When
$sign(t_4) = sign(t'_b)$, $I_{-N}^2 > I_{N}^2$, so that $\chi_0({\bf
Q}_{-N} ) > \chi_0({\bf Q}_{N} )$ and a phase with  {\it negative even}
$N$ is favored for some values of the field.

Fig. \ref{fig4} shows the evolution of the maximum of the susceptibility
with the
field. Is is made of different sheets corresponding to the different nesting
vectors. It is seen that
 for the standard model ($t_3=t_4=0$), all the SDW phases will have a
positive $N$, and that for some finite $t_3$ and $t_4$, negative
phases appear and the sequence $1,2,-2,3,4,-4,5$ is found, as observed
 experimentally\cite{Cooper89}.


Fig. \ref{fig3}  shows the  variation of the Hall voltage with the field. It
is qualitatively  similar  to the experimental
ones\cite{Cooper89,Piveteau86,Balicas95}. At low $T$, a sequence of fine
quantized  structures may be
resolved and that an alternance of many subphases
with positive and negative quantum numbers can appear.
It is in very good agreement with the complex structure
observed  ten years ago, near  the threshold field, see fig.\ref{fig5}
\cite{Piveteau86}.

\begin{figure}[htb]
\centerline{
\epsfysize 11cm
\epsffile{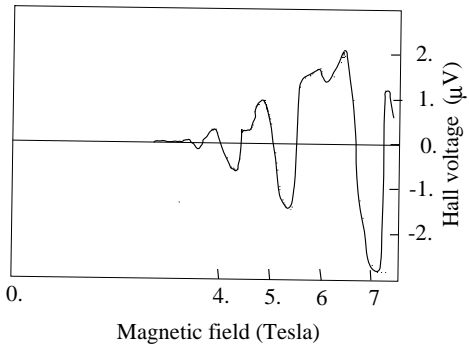}}
\caption{ \footnotesize{The Hall voltage in $(TMTSF)_2 PF_6$ under
pressure, from \protect\cite{Piveteau86}. It is qualitatively well
described by the result of fig. \ref{fig3}. } } \label{fig5}
\end{figure}

\begin{figure}[bht]
\centerline{
\epsfysize 10cm
\epsffile{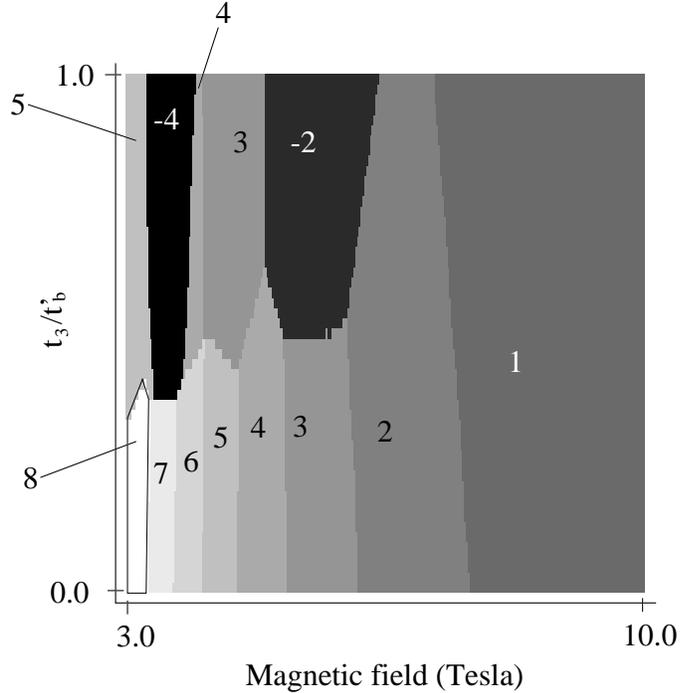}}
\caption{\footnotesize{Phase diagram showing the Hall numbers vs. field
for differents values of the parameter
 $t_3/t'_b$. Here $t'_b=10K$. The field scale is proportional to $t'_b$.} }
\label{fig6}
\end{figure}

 We believe that the
experimentally observed high sensitivity of the sequence of sign reversal
to external pressure is due to the sensitivity of the parameters of the
dispersion relation to pressure. We emphasize that although
 the metallic electron gas  has a metallic behavior with a large Fermi energy,
 of the order of the $eV$, the cascade of SDW subphases
 is driven by extremely small energy scales, of the order of a few Kelvins.
 The orders of magnitude of these two additional harmonics are non
incompatible
with the estimations of  a refined microscopic  model\cite{Yamaji86}

Finally it is worth noticing that positive and negative phases are almost
degenerate because the energy scale $t_4$ is certainly very small, of the
order of $1K$. The relative energy difference between these two phases
is very small, of order $(t_4/\omega_c)^2$. This explains which the negative
phases are always very
sensitive to external parameters like pressure or probably anion ordering.

Fig. \ref{fig6} shows the evolution of the QHE sequence with the parameter
$t_3$. One sees that the even negative phases appear when $t_3$ increases.

\bigskip

{\bf 5. Conclusion}
\medskip

The Quantized Nesting Model explains very well the quantization of the Hall
effect observed in Bechgaard salts. The observed structure is extremely
sensitive to the details of the dispersion relation and the nesting of the
Fermi surface. We have explained the ten-years-old puzzle of the observed
sign reversals of the Quantum Hall effect in the cascade of FISDW phases of
Quasi-1D organic conductors. They can be described
 with a slight modification of the dispersion relation of the
metallic phase.
We have been able to reproduce the observed sequence of "negative" phases
with an even quantum number, to understand why they are very sensitive to
pressure and why it is more difficult to measure a well defined plateau.
  Our result shows that the electronic properties of the
Bechgaard salts are extremely sensitive to very small changes in the
geometry of the FS and that the standard model and its variations still
continues to describe very well the observed phase diagram of these salts in
a magnetic field.

Acknowledgments:   We thank  L. Balicas, A. Bjeli\u{s}, M. H\'eritier, D.
J\'erome,  G. Kriza and  M.
Ribault for very useful discussions.
This work was supported in part by NATO Grant No 890191.


}

\end{document}